\documentclass[prd,floatfix,onecolumn,amsmath,amssymb]{revtex4-1}
\usepackage{graphicx,color,dcolumn,booktabs,bm}
\usepackage{subfigure}
\usepackage{amssymb}
\usepackage{longtable}
\usepackage{indentfirst}
\usepackage{epsfig,gensymb,siunitx}
\usepackage{epstopdf}   %{feynmp}
\usepackage{slashed}  %for Feynman symbols
\usepackage{cases}
\usepackage[pdfpages]{xcolor}
\definecolor{maroon}{RGB}{139,25,150}%burada 0-255 arasi her biri icin numara vererek renk elde et
\usepackage{multirow}
\usepackage{float}
\usepackage{pgf}
\usepackage{physics}
\usepackage[colorlinks, citecolor=blue,anchorcolor=red,menucolor=red, linkcolor=red,filecolor=red,runcolor=red,urlcolor=blue,frenchlinks=red, urlcolor=blue]{hyperref}

\usepackage{tikz}
\usetikzlibrary{decorations.pathmorphing}

\begin{document}

	\preprint{}
	
\title{\color{maroon}{Determination of proton electromagnetic form factors from  DVCS measurements}}

	\author{The MMGPDs Collaboration:\\
Anoushiravan Moradi$^{1}$}
%\email{a.moradi1992@ut.ac.ir}
	
\author{Muhammad Goharipour$^{2,3}$}
%\email{muhammad.goharipour@ipm.ir}

\author{H.~Fatehi$^{1}$}
%\email{apranik.fatehi@cern.ch}

\author{K.~Azizi$^{1,4,3}$}
\email{kazem.azizi@ut.ac.ir}
\thanks{Corresponding author}

%\author{\\ And\\ Kornelija Passek-K.$^{5}$}
%\email{passek@irb.hr}
	
\affiliation{
$^{1}$Department of Physics, University of Tehran, North Karegar Avenue, Tehran 14395-547, Iran\\
$^{2}$School of Physics, Institute for Research in Fundamental Sciences (IPM), P.O. Box  19395-5531, Tehran, Iran\\
$^{3}$School of Particles and Accelerators, Institute for Research in Fundamental Sciences (IPM), P.O. Box 19395-5746, Tehran, Iran\\
$^{4}$Department of Physics, Dogus University, Dudullu-\"{U}mraniye, 34775 Istanbul, T\"urkiye }
%$^{5}$Division of Theoretical Physics, Rudjer Bo\v{s}kovi\'{c} Institute, HR-10000 Zagreb, Croatia}
	
\date{\today}

\begin{abstract}

We present a detailed analysis of the proton electromagnetic form factors (FFs) using exclusive photon leptoproduction (EP) data in kinematic regions where the Bethe–Heitler (BH) contribution dominates the deeply virtual Compton scattering (DVCS) cross section.
By exploiting the sensitivity of the BH amplitude to the Dirac and Pauli FFs, we extract $F_{1}(t)$, $F_{2}(t)$, and the corresponding Sachs FFs within several fitting scenarios based on dipole and $P$-pole parametrizations, and evaluate the charge and magnetic radii of the proton.
In this fitting scenario, we show that EP measurements in the range $0.11 < |t| < 0.45~\mathrm{GeV}^2$ can provide constraints on $F_1(t)$, while offering limited sensitivity to $F_2(t)$.
The extracted charge radius values tend to be smaller than those obtained from traditional elastic electron–proton scattering measurements and are consistent, within uncertainties, with recent high-precision PRad results. These findings indicate that EP measurements, especially when covering smaller values of $|t|$, can serve as a complementary tool for determining the proton electromagnetic structure and may contribute to ongoing efforts to better understand the proton charge radius.
The methodology developed here provides a framework for future combined analyses of EP and elastic electron-proton scattering data which enables a unified determination of the nucleon FFs.

\end{abstract}

\maketitle

%%%%%%%%%%%%%%%%%%%%%%%%%%%%%%%%%%%%%%%%%%%%%%%%%%%%%%%%%%%%%%%%%%%%%%%%%%%%%%%%%%%
\section{Introduction}\label{sec:one} 

It has been well stabilized now that deeply virtual Compton scattering (DVCS), an exclusive process where a lepton (e.g., electron) scatters off a hadron (e.g., proton) by exchanging a virtual photon and a real photon is produced in the final state, leaving the target intact, is  one of the most powerful tools for probing the internal structure of hadrons~\cite{Ji:1996nm,Belitsky:2001ns}.
In contrast to inclusive deep inelastic scattering (DIS), which provides only information on longitudinal momentum distributions of partons through the parton distribution functions (PDFs)~\cite{Alekhin:2002fv}, DVCS provides access to generalized parton distributions (GPDs)~\cite{Ji:1996ek,Ji:1996nm,Radyushkin:1996nd,Radyushkin:1997ki,Collins:1998be,Belitsky:2001ns,Guidal:2013rya,Kumericki:2016ehc,Aschenauer:2025cdq,Guo:2025muf,Xie:2023xkz,Braun:2025noa}. GPDs are more comprehensive objects compared to PDFs and represent the correlations between the spatial and momentum degrees of freedom inside the nucleon~\cite{Diehl:2003ny,Ji:2004gf,Belitsky:2005qn,Boffi:2007yc,Diehl:2015uka,Kriesten:2020wcx,Mezrag:2022pqk,Lorce:2025aqp,Boer:2025ixc}. Through these distributions, DVCS can provide insights into fundamental aspects of quantum chromodynamics (QCD), such as the three-dimensional imaging of the nucleon through GPDs~\cite{Burkardt:2002hr,Kaur:2023lun,Cichy:2024afd}, the decomposition of the nucleon spin among its quark and gluon constituents~\cite{Aidala:2012mv,Ji:2020ena}, the connection between various nucleon form factors (FFs) and partonic correlations~\cite{Diehl:2003ny}, and the mechanical properties of the nucleon~\cite{Polyakov:2018zvc,Burkert:2018bqq,Kumericki:2019ddg,Martinez-Fernandez:2025jvk}.

The theoretical framework of QCD factorization ensures that the DVCS amplitude can be expressed in terms of perturbatively calculable coefficient functions and universal GPDs, which enter through the so-called Compton form factors (CFFs)~\cite{Collins:1998be,Guo:2023ahv,CaleroDiaz:2025luc}.
Experimentally, DVCS has been investigated extensively at HERA, JLab, COMPASS, and HERMES, leading to a wealth of precise measurements of cross sections and related observables such as beam-spin, beam-charge, and target-spin asymmetries (see Ref.~\cite{Kumericki:2016ehc} and references therein). These observables are highly sensitive to the underlying GPDs. Therefore, the precise measurements of DVCS observables can be used in global analyses in order to extract the functional form of GPDs~\cite{Kumericki:2016ehc,Kriesten:2019jep,Guo:2023ahv}.
Moreover, measurements of DVCS necessarily include the Bethe-Heitler (BH) process, in which the real photon is emitted from the lepton rather than the nucleon,
and which often dominates the cross section in the experimentally accessible kinematic regions. The interference between the DVCS and BH amplitudes plays a central role in unraveling the nucleon structure~\cite{Kriesten:2019jep}.

GPDs are directly linked, through their Mellin moments, to a wide class of hadronic FFs, including electromagnetic~\cite{Guidal:2004nd,Diehl:2013xca,Hashamipour:2022noy}, axial~\cite{Hashamipour:2019pgy,Irani:2023lol,Bhattacharya:2024wtg}, gravitational~\cite{Pagels:1966zza,Polyakov:2018zvc,Goharipour:2025lep}, and transition FFs~\cite{Goharipour:2024atx}. Among these, the electromagnetic FFs have a central role in hadron structure studies. They are commonly expressed either in terms of the electric and magnetic Sachs
form factors, $G_E(t)$ and $G_M(t)$, or equivalently in terms of the Dirac and Pauli form factors, $F_1(t)$ and $F_2(t)$, where $t$ is the Mandelstam variable
corresponding to the squared four-momentum transfer to the nucleon. In the kinematic region considered in this work, $t$ is negative.
 These quantities encode the spatial distributions of electric charge and magnetic moment within the nucleon. In particular, precise measurements of these FFs allow one to determine the nucleon charge and magnetic radii, $r_E$ and $r_M$, which are directly related to the slopes of the FFs at $t=0$~\cite{Karr:2020wgh,Gao:2021sml,Xiong:2023zih}. Such measurements have attracted great interest in recent years, due to the so-called proton charge radius puzzle~\cite{Lorenz:2012tm,Pohl:2013yb,Bernauer:2014cwa,Carlson:2015jba,Hill:2017wzi,Peset:2021iul,Lin:2023fhr,Lumpay:2025btu}, which refers to the discrepancy observed in measurements of the proton's charge radius when using different methods. Beyond their static interpretation, electromagnetic FFs can also be considered as important benchmarks for testing various theoretical and phenomenological approaches to QCD, such as lattice simulations~\cite{Park:2021ypf}, holographic QCD~\cite{Xu:2021wwj}, and phenomenological GPD models~\cite{Goharipour:2024mbk} (see Ref.~\cite{Goharipour:2025yxm} and references therein for more information). Consequently, the extraction of $F_1$ and $F_2$ from experimental observables remains an important mission in hadronic physics.

To extract the electromagnetic FFs from the experimental data, the measurements of the elastic electron-nucleon scattering cross section is usually used~\cite{Gao:2003ag,Qattan:2004ht,Perdrisat:2006hj,Crawford:2006rz,Arrington:2007ux,A1:2010nsl,A1:2013fsc,Ye:2017gyb,Xiong:2019umf,Christy:2021snt,Williams:2025fiv,Vaziri:2025rfq}. In the present study, we aim to examine the possible extraction of these FFs from the measurements of the exclusive photon leptoproduction (EP) process which comprises the DVCS and BH amplitudes as well as their interference. To be more precise, since the BH contribution to the EP process is directly sensitive to the electromagnetic FFs, a novel idea to extract them is to analyse EP data in kinematic regions where the BH term is strongly dominant over the other contributions. To this aim, we first examine this idea by calculating different contributions in the EP cross section separately for a nominal energy scale $ Q^2=1 $ GeV$ ^2 $ and searching for the kinematic regions where the BH contribution is dominant. Then, we extract Dirac and Pauli FFs, $ F_1(t) $ and $ F_2(t) $, by analyzing the JLab data considering only data points for which the BH contribution is dominant.

%%%%%%%%%%%%%%%%%%%%%%%%%%%%%%%%%%%%%%%%%%%%%%%%%%%%%%%%%%%%%%%%%%%%%%%%%
\section{EP cross section and BH contribution}\label{sec:two} 

\begin{figure}[!htb]
    \centering
\includegraphics[scale=0.3]{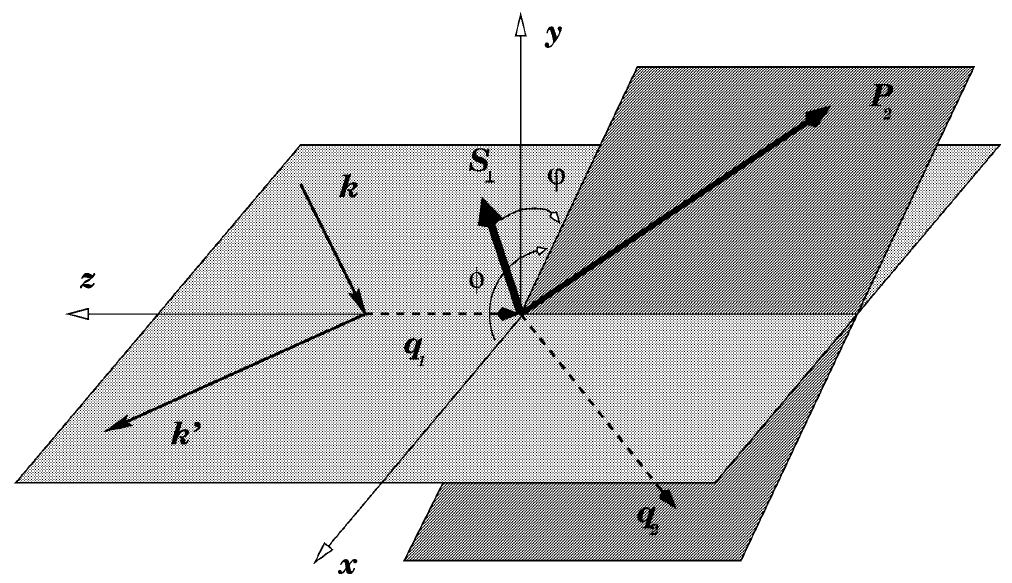}   
\caption{EP process shown in terms of laboratory-frame kinematic variables. Figure taken from Ref.~\cite{Belitsky:2001ns}.}
\label{fig:DVCS}
\end{figure}

One of the most insightful approaches to access GPDs is through exclusive processes~\cite{Ji:1998pc,Goeke:2001tz,Kroll:2012sm,Duplancic:2022ffo,Qiu:2022pla}. Exclusive processes are characterized by the requirement that the hadron remains intact in the final state. Among the most important of these processes are DVCS~\cite{Ji:1996ek,Ji:1996nm,Radyushkin:1996nd,Radyushkin:1997ki,Collins:1998be,Belitsky:2001ns,Guidal:2013rya,Kumericki:2016ehc,Aschenauer:2025cdq,Guo:2025muf,Xie:2023xkz,Braun:2025noa} and deeply virtual meson production (DVMP)~\cite{Collins:1996fb,Goloskokov:2005sd,Goloskokov:2007nt,Muller:2013jur,Favart:2015umi,Cuic:2023mki,Boussarie:2024pax,Hatta:2025vhs}. In both cases, the target remains intact, while an additional particle is produced: a real photon in DVCS and a meson in DVMP. In the present study, we focus on EP, which is the experimentally accessible process underlying both DVCS and BH mechanisms.
The EP process is described by
\begin{equation}
	e(k) + p(P_1) \to e(k') + p(P_2) + \gamma(q_2)\,,
\label{Eq1}	
\end{equation}
in which an incoming lepton with four-momentum $k$ scatters off a proton of momentum $P_1$, producing an outgoing lepton $k'$, a recoiling proton $P_2$, and a real photon $q_2$ (see Fig.~\ref{fig:DVCS}). 

The first computation of the DVCS cross section in terms of leading-twist CFFs was carried out by Ji~\cite{Ji:1996nm}. Subsequently, Belitsky, Müller, and Kirchner (BKM)~\cite{Belitsky:2001ns} made a substantial contribution by extending the analysis beyond the leading-order treatment of GPDs. They provided explicit formulas for DVCS cross section in terms of laboratory-frame kinematic variables, as illustrated in Fig.~\ref{fig:DVCS}. In their convention, the azimuthal angle $\phi$ denotes the orientation of the recoiling hadron. Henceforth, we adopt the notation $\phi$ to denote the azimuthal angle of the recoiling hadron. Other relevant kinematic quantities, such as the polar and azimuthal angles of the lepton, which are not explicitly shown in Fig.~\ref{fig:DVCS}, are defined below as needed. A convenient reference frame is chosen such that the virtual photon four-momentum $q_1$ has no transverse components. The $z$-axis is fixed along the negative direction of $q_1$, while the incoming lepton has a positive $x$-component. Explicitly, the relevant four-momenta are defined as
\begin{align}
k   &=  \left(E,\, E \sin\theta_\ell,\, 0,\, E \cos\theta_\ell \right)\,, \nonumber \\
q_1 &= \left(q_1^{0},\, 0,\, 0,\, -|q_1^{3}| \right)\,, \nonumber \\
P_1 &= \left(M,\, 0,\, 0,\, 0 \right)\,, \nonumber \\
P_2 &= \left(E_2,\, |P_2| \cos\phi \sin\theta_N,\, |P_2| \sin\phi \sin\theta_N,\, |P_2| \cos\theta_N \right)\,.
\label{Eq2}	
\end{align}
In this frame, $\theta_\ell$ denotes the polar angle of the incoming lepton with respect to the negative $z$-axis (the virtual photon direction), while $\theta_N$ represents the polar angle of the recoiling proton relative to the same axis. The azimuthal angle of the scattered lepton is fixed to $\phi_\ell = 0$, while, as defined in Fig.~\ref{fig:DVCS}, the angle between the lepton plane and the recoiling proton momentum is given by $\phi$. Here $E$ is the energy of the incoming lepton, $E_2$ the energy of the recoiling proton, and $M$ the proton mass. Finally, the longitudinal polarization vector is taken as $S_{LP} = \left(0,\, 0,\, 0,\, \Lambda \right)$. $ \varphi $ represents the angle between this vector and the hadronic plane.

\begin{figure}[!htb]
    \centering
\includegraphics[scale=0.45]{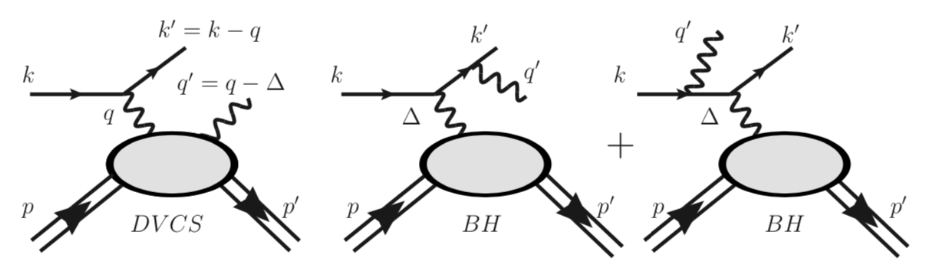}   
\caption{Diagrammatic representation of DVCS and BH processes. In DVCS (left), the real photon is emitted by the struck quark inside the proton, whereas in BH it is emitted from the lepton line, either after (center) or before (right) scattering off the proton. Figure taken from Ref.~\cite{Adams:2024pxw}.}
\label{fig:BH}
\end{figure}

In addition to DVCS, there is another process with identical initial and final states, which cannot be experimentally distinguished, known as the BH process~\cite{Belitsky:2001ns}. Figure~\ref{fig:BH} illustrates the DVCS diagram (left) together with the two BH diagrams (center and right). In the BH case, the incoming electron is elastically scattered by the nucleon, while the real photon is emitted from the lepton line rather than from the struck quark, either after (center) or before (right) the lepton scatters off the nucleon. The four-momenta are assigned as $q \equiv q_1$ for the virtual photon, $q' \equiv q_2$ for the real photon, and $\Delta = p' - p = q- q' $ with $t = \Delta^2$, where $p \equiv P_1$ and $p' \equiv P_2$. Unlike DVCS, the BH process is entirely determined by the elastic nucleon FFs, $F_1(t)$ and $F_2(t)$~\cite{Belitsky:2001ns}.

At leading order in the electromagnetic coupling $ \alpha_{\text{QED}} $, the scattering amplitude receives contributions from both the DVCS and BH processes. These two processes cannot be experimentally distinguished, and therefore the total amplitude is given by their coherent sum. Consequently, the squared amplitude that determines the EP cross section contains three contributions: the pure BH term, the pure DVCS term, and an interference term $ \mathcal{I}  $ arising from the combination of the two amplitudes,
\begin{equation}
  |\mathcal{T}|^2 
    = |\mathcal{T}_{\text{BH}}|^2 + |\mathcal{T}_{\text{DVCS}}|^2 + 
    \underbrace{ \mathcal{T}_{\text{BH}} \, \mathcal{T}_{\text{DVCS}}^\ast 
    + \mathcal{T}_{\text{BH}}^\ast \, \mathcal{T}_{\text{DVCS}} }_{\text{Interference} ~ \mathcal{I}}\,,
\label{Eq3}
\end{equation}
where the symbol ${}^\ast$ denotes complex conjugation.

The four-fold differential cross section of EP can be expressed as~\cite{Belitsky:2010jw,Benali:2020vma}:
\begin{equation}
	\frac{d^4\sigma}{dQ^{2}\, dx_{B}\, dt\, d\phi}
	= \frac{\alpha_{\text{QED}}^{3} x_{B} y^{2}}{8\pi e^6 Q^{4} \sqrt{1+\epsilon^{2}}}
	\left(
	\left| \mathcal{T}_{\text{BH}} \right|^{2}
	+ \left| \mathcal{T}_{\text{DVCS}} \right|^{2}
	+ \mathcal{I}
	\right)\,,
	\label{Eq4}
\end{equation}
where $\epsilon = 2 x_B M/Q$, with $x_B = Q^2/(2 p \cdot q)$ and $Q^2 = -q^2$, $e$ is the elementary charge, and $\alpha_{\text{QED}}$ is the electromagnetic fine-structure constant. The variable $y$ denotes the fractional energy loss of the lepton, defined in the laboratory frame as
\begin{equation}
y = \frac{q \cdot p}{k \cdot p} \,.
\label{Eq5}
\end{equation}
Within the BKM framework, the DVCS, BH, and interference contributions are expressed through a harmonic expansion in $\phi$, extended up to twist-3 accuracy and incorporating gluon transversity effects. 
In the following, we focus on the BH term.

The BH amplitude, calculated within the framework of pure QED, can be expressed in terms of the nucleon elastic FFs. 
The squared amplitude of BH is given by the following expression~\cite{Belitsky:2001ns}:
\begin{equation}
	\left| \mathcal{T}_{\text{BH}} \right|^{2} =
	\frac{e^6}{x_{B}^{2} y^{2} (1 + \varepsilon^{2})^{2}\, t \, P_{1}(\phi)\, P_{2}(\phi)}
	\left[
	c_{0}^{\text{BH}} +
	\sum_{n=1}^{2} c_{n}^{\text{BH}} \cos(n\phi) +
	s_{1}^{\text{BH}} \sin(\phi)
	\right]\,,
	\label{Eq6}
\end{equation}
where \(P_{1}(\phi)\) and \(P_{2}(\phi)\) denote the lepton propagators corresponding to the two BH diagrams. The complete expressions for these propagators and the harmonic coefficients \(c_{i}^{\text{BH}}\) and \(s_{1}^{\text{BH}}\) are provided in Ref.~\cite{Belitsky:2001ns}. 
For an unpolarized target, the sine coefficient vanishes, i.e., \(s_{1}^{\text{BH}} = 0\). The harmonic coefficients are functions of the proton electromagnetic FFs \(F_1(t)\) and \(F_2(t)\).
This dependence provides a unique opportunity: in kinematic regions where the BH contribution dominates over the pure DVCS and interference terms, the measured cross section becomes directly sensitive to the proton's electromagnetic FFs. 
This observation underlies the central motivation of the present study.

The explicit dependence of the BH cross section on the electromagnetic FFs becomes particularly transparent in an alternative formalism proposed in Refs.~\cite{Kriesten:2019jep}, where the BH contribution is cast in a Rosenbluth-type representation, analogous to that used in elastic lepton-nucleon scattering. In this framework, the unpolarized BH cross section naturally separates into terms proportional to the Dirac and Pauli FFs, that makes the individual roles of $F_{1}$ and $F_{2}$ explicit. This decomposition provides direct sensitivity to the underlying elastic structure of the nucleon even within the deeply virtual kinematic regime. The unpolarized BH contribution to the four-fold differential cross section can be written as~\cite{Kriesten:2019jep,Adams:2024pxw}:
\begin{equation}
	\frac{d^{4}\sigma^{\text{BH}}_{\text{unpol}}}{dx_{B}\, dQ^{2}\, d|t|\, d\phi}
	= \frac{\Gamma}{t}
	\left\{ 
	A_{\text{BH}} \left[ F_1^2(t) + \tau F_2^2(t) \right] 
	+ B_{\text{BH}}\, \tau \left[ F_1(t) + F_2(t) \right]^2 
	\right\},
	\label{Eq7}
\end{equation}
where $\tau = -t/(4M^{2})$, and
\begin{equation}
	\Gamma = \frac{\alpha_{\text{QED}}^3}{8\pi (s - M^2)^2 \, \sqrt{1 + \gamma^2} \, x_B}\,,
	\label{Eq8}
\end{equation}
with $\gamma = 4M^2 x_B^2/Q^2$ and $s = (k + p)^2$ denoting the electron-proton center-of-mass energy squared. Note that the Dirac and Pauli FFs are related to the Sachs FFs as $ G_E= F_1 - \tau F_2 $ and $ G_M= F_1 + F_2 $.
The coefficients \(A_{\text{BH}}\) and \(B_{\text{BH}}\) are dimensionless kinematic functions of \((s, Q^2, x_B, t, \phi)\). Their explicit expressions were derived in Ref.~\cite{Kriesten:2019jep}, where they are presented in a fully covariant form. Equation~(\ref{Eq7}) thus demonstrates how the BH term can be organized in a Rosenbluth-like framework, providing a transparent and model-independent link between the measurable cross section and the elastic nucleon FFs.

%%%%%%%%%%%%%%%%%%%%%%%%%%%%%%%%%%%%%%%%%%%%%%%%%%%%%%%%%%%%%%%%%%%%%%%%%
\section{Searching for dominant BH}\label{sec:three}

In the previous section, we showed that the BH amplitude is directly sensitive to the Dirac and Pauli FFs, $F_1$ and $F_2$, and hence to the Sachs form factors, $G_E$ and $G_M$. This raises the question of whether there exist kinematic regions in which the BH contribution dominates the DVCS cross section, so that the electromagnetic FFs can be extracted through the analysis of EP data. In this section, we investigate this possibility by exploring the dependence of the cross section on two relevant variables $x_B$ and $\phi$, at fixed kinematics $Q^2 = 1~\text{GeV}^2$, $t = -0.1~\text{GeV}^2$, and $ E=6 $ GeV, as a representative example. To this aim, we utilize the \texttt{Gepard} package~\cite{Kumericki:2006xx,Kumericki:2007sa,Kumericki:2009uq,Cuic:2023mki}, which provides the capability to compute the individual contributions to the EP cross section. In addition, the KM15 model~\cite{Kumericki:2015lhb}, as implemented in \texttt{Gepard}, is used to perform the numerical calculations presented in this section.

\begin{figure}[!htb]
    \centering
\includegraphics[scale=0.78]{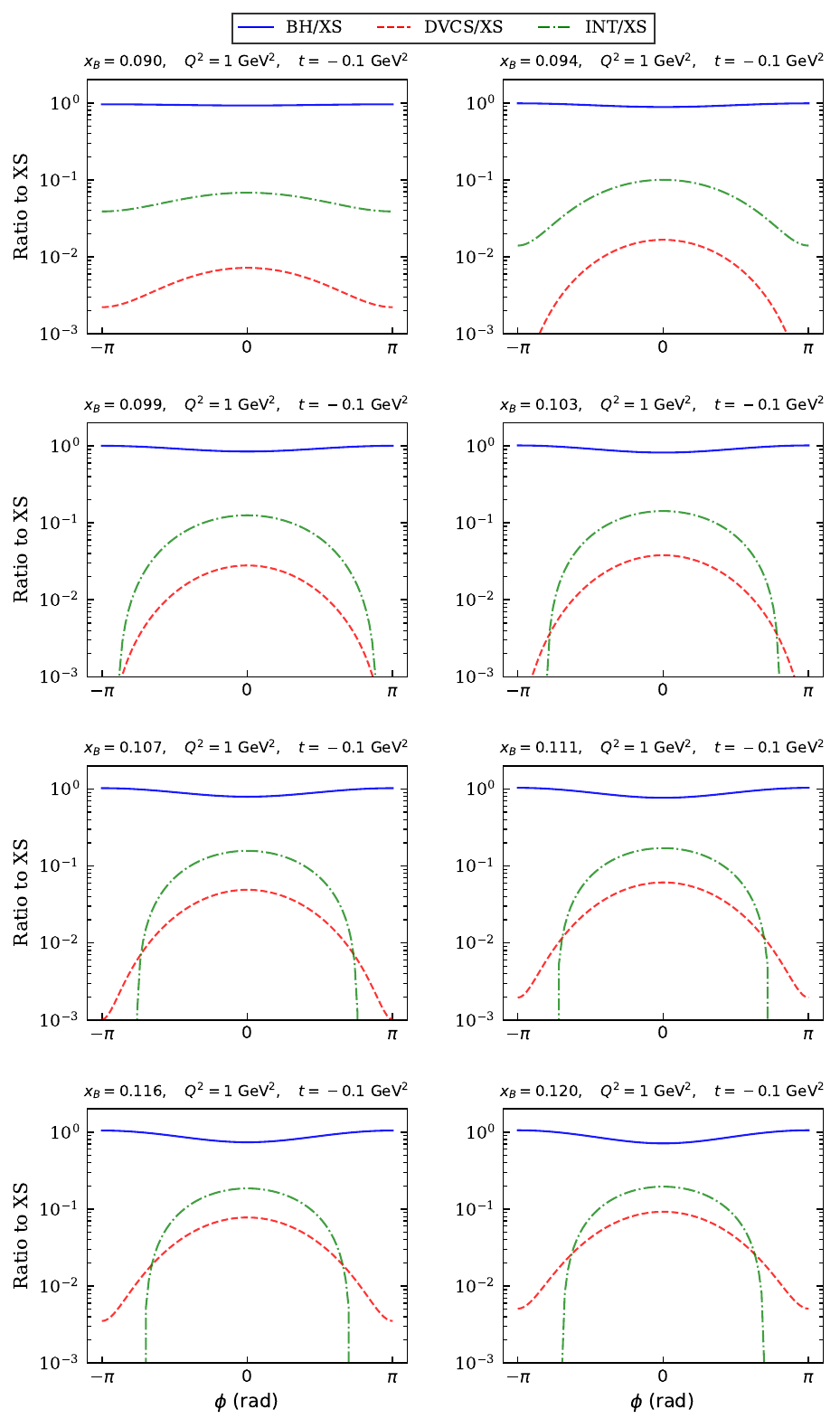}   
\caption{The ratio of the BH, DVCS, and $ \mathcal{I} $ (denoted as INT) contributions to the four-fold EP cross section of Eq.~(\ref{Eq4}) (denoted as XS), shown separately as functions of $\phi$. The results are calculated using the \texttt{Gepard} package~\cite{Kumericki:2006xx,Kumericki:2007sa,Kumericki:2009uq,Cuic:2023mki} with the KM15~\cite{Kumericki:2015lhb} model at fixed kinematics $Q^2 = 1~\text{GeV}^2$ and $t = -0.1~\text{GeV}^2$, for different values of $x_B$.}
\label{fig:XSContributions}
\end{figure}

Figure~\ref{fig:XSContributions} shows the results obtained for the ratio of the
BH, DVCS, and $ \mathcal{I} $ (denoted as INT) contributions to the four-fold EP cross section of Eq.~(\ref{Eq4}) (denoted as XS). Each panel corresponds to a fixed value of $x_B$, and the results have been plotted as a function of the $\phi$. As can be seen, in the considered kinematic region of low $Q^2$ and $|t|$, the BH contribution is significantly larger than those from DVCS and INT, as expected. This behavior is consistent with the fact that the BH process is purely electromagnetic and becomes suppressed with increasing $Q^2$ and $|t|$.

Examining the $x_B$ dependence more precisely, one can realize a non-trivial angular behavior of the DVCS and INT terms.
As $x_B$ increases, the magnitude of the INT contribution decreases at large values of $\phi$ but increases at small values of $\phi$. The situation is somewhat different for DVCS term. Its contribution decreases at large $\phi$ for moderate $x_B$ but begins to rise again at higher $x_B$. At small $\phi$, DVCS behaves like the INT term and grows steadily with increasing $x_B$
According to these results, the BH contribution remains significantly dominant at large $\phi$. For $x_B \geqslant 0.099$, one can clearly identify regions where the BH contribution accounts for more than 95\%. In some cases, more than 99\% of the total EP cross section is provided by the BH contribution. 

These results indicate that, at $Q^2 = 1~\text{GeV}^2$ and $t = -0.1~\text{GeV}^2$, there exist kinematic regions in $ x_B $ and $ \phi $ where the EP cross section is almost entirely reproduced by the BH contribution. In fact, for any fixed values of $Q^2$ and $|t|$, one can find similar regions where BH dominance holds. In general, as $Q^2$ and $|t|$ increase, the $x_B$ interval where BH dominance occurs shifts toward larger $x_B$ values.  
These observations suggest that one may impose suitable cuts on the available experimental EP data in order to isolate those data points for which the BH term dominates. Such a selection would allow the extraction of the Dirac and Pauli form factors, $F_1(t)$ and $F_2(t)$, by performing a $\chi^2$ analysis with appropriate parametrization forms. This approach provides an alternative method for determining $F_1(t)$ and $F_2(t)$, complementary to the traditional extraction from elastic electron-nucleon scattering cross sections. It is of particular interest to compare the results of these two approaches and examine possible differences in the extracted FFs.

%
%%%%%%%%%%%%%%%%%%%%%%%%%%%%%%%%%%%%%%%%%%%%%%%%%%%%%%%%%%%%%%%%%%%%%%%%%

\section{Data selection}\label{sec:four}

As mentioned earlier, our goal is to extract the Dirac and Pauli form factors, $F_{1}(t)$ and $F_{2}(t)$ through a $\chi^{2}$ analysis of EP cross section data. For this purpose, we use the four-fold differential cross sections of Eq.~(\ref{Eq4}) measured by the CLAS Collaboration for the $ep \to e' p' \gamma$ reaction using the CLAS detector and the 5.75-GeV electron beam at Jefferson Lab~\cite{CLAS:2015uuo}. These measurements cover a wide kinematic range in the variables $(Q^2, x_B, t, \phi)$.

However, a substantial fraction of the CLAS data lies in kinematic regions where the BH contribution is not dominant, such that the pure BH approximation cannot be reliably applied. 
Note also that the experimental EP measurements do not separate the BH, DVCS, and interference contributions. So, it is not possible to determine from the data alone whether a specific point lies in
a BH-dominated kinematic region. To isolate the region in which the BH term provides an accurate description of the cross section, we construct five reduced datasets derived from the full CLAS measurements. Each data set includes only those points for which the relative difference between the theoretical prediction using the BH contribution alone and the full EP cross section is less than 1\%, 2\%, 3\%, 4\%, or 5\%, respectively. We use the Gepard package~\cite{Kumericki:2006xx,Kumericki:2007sa,Kumericki:2009uq,Cuic:2023mki} in order to evaluate the ratio $ R_{\mathrm{BH}} = \frac{\sigma_{\mathrm{BH}}}{\sigma_{\mathrm{EP}}} $.

Table~\ref{tab:Data} summarizes these five datasets, listing the number of data points (N$_{\rm pts.}$) included in each case. All sets span the ranges $ 0.110 < |t| < 0.450 $ GeV$ ^2 $ and $ -3.012 \lesssim  \phi \lesssim 3.012 $ deg. Except for Set~1, which covers $ 0.184 < x_B < 0.475 $ and $ 1.450 < Q^2 < 3.770 $ GeV$ ^2 $, the remaining sets span a slightly wider kinematic region, $ 0.154 < x_B < 0.475 $ and $ 1.270 < Q^2 < 3.770 $ GeV$ ^2 $.
\begin{table}[th!]
\centering
\caption{Summary of the five datasets constructed from the CLAS EP measurements~\cite{CLAS:2015uuo}. Each data set includes only those points for which the relative difference between the EP cross section and the pure BH contribution is below a chosen threshold. The table lists the BH selection criterion and the number of retained data points.}
\label{tab:Data}

\begin{tabular}{c
        @{\hspace{1.2cm}} c
        @{\hspace{1.2cm}} c}
\hline\hline
\textbf{Dataset} & \textbf{BH criterion} & \textbf{N$_{\rm pts.}$} \\
\hline
Set 1 & $\Delta\sigma/\sigma \le 1\%$ & 98  \\[3pt]
Set 2 & $\Delta\sigma/\sigma \le 2\%$ & 212 \\[3pt]
Set 3 & $\Delta\sigma/\sigma \le 3\%$ & 292 \\[3pt]
Set 4 & $\Delta\sigma/\sigma \le 4\%$ & 404 \\[3pt]
Set 5 & $\Delta\sigma/\sigma \le 5\%$ & 473 \\
\hline\hline
\end{tabular}
\end{table}

As expected, the most inclusive data set which corresponds to the 5\% criterion (where the BH contribution accounts for at least 95\% of the full cross section) contains the largest number of points. To account for uncertainties associated with this selection procedure, we assign an additional uncorrelated systematic uncertainty of 1–5\% to the retained data points, matched to the dominant-BH criterion used. However, we note that our analysis indicates that the final extracted results for $F_{1}(t)$ and $F_{2}(t)$ are not significantly affected by this additional uncertainty.
We also note that the present analysis uses the published cross-section data as reported by the CLAS collaboration. No additional unfolding or bin-migration corrections are applied.

%
%%%%%%%%%%%%%%%%%%%%%%%%%%%%%%%%%%%%%%%%%%%%%%%%%%%%%%%%%%%%%%%%%%%%%%%%%

\section{Results}\label{sec:five} 

In this section, we present the results obtained from the analysis of the five datasets introduced in the previous section. Using a dipole parametrization for the Dirac and Pauli form factors, $F_{1}(t)$ and $F_{2}(t)$, we first examine the differences between the five fits. We also investigate the impact of replacing the dipole form with a more flexible $P$-pole parametrization to test the stability of the extracted FFs. Furthermore, considering the fact that the EP data used in this work do not provide strong constraints on the Pauli form factor $F_{2}(t)$, we perform an additional analysis in which $F_{2}(t)$ is fixed to the Kelly's parametrization~\cite{Kelly:2004hm}, allowing only $F_{1}(t)$ to vary in the fit.

To perform the $\chi^{2}$ minimization and extract the optimal values of free parameters along with their uncertainties, we use the Python interface of the CERN \texttt{MINUIT} library~\cite{James:1975dr}, \texttt{iMinuit}~\cite{iminuit}. The uncertainties on the extracted FFs are estimated using the standard Hessian method as implemented in \texttt{iMinuit}, with $\Delta\chi^{2} = 1$ corresponding to the $68\%$ confidence level.

\subsection{Dipole fit}

Since the data included in our analysis do not cover a wide range in $|t|$ and are restricted to the small-$|t|$ region ($0.110 < |t| < 0.450~\mathrm{GeV}^2$; see Sec.~\ref{sec:four} for further information), we first consider a simple dipole parametrization for the Dirac and Pauli form factors:
\begin{align}
F_1(t) &= \left(1 - \frac{t}{a_E}\right)^{-2}, \nonumber \\
F_2(t) &= \kappa \left(1 - \frac{t}{a_M}\right)^{-2}, 
\label{Eq9}
\end{align}
where $\kappa = 1.793$ is the anomalous magnetic moment of the proton in units of the nuclear magneton, and $a_E$ and $a_M$ are free parameters to be determined from the fit.

Table~\ref{tab:chi2} summarizes the values of $\chi^{2}$ divided by the number of degrees of freedom, $\chi^{2}/\mathrm{d.o.f.}$, for each fit. As can be seen, the lowest (and thus best) values correspond to the analyses with the largest number of data points (Fit~4 and Fit~5). It is important to recall that as the BH-dominance criterion becomes less restrictive, the additional systematic uncertainty assigned to the data also increases (from 1\% to 5\%). This effect, together with the larger number of data points included in the fits, contributes to the systematic decrease in $\chi^{2}/\mathrm{d.o.f.}$ from Fit~1 to Fit~5. Table~\ref{tab:fit_params} shows the optimal values of the fit parameters, $a_E$ and $a_M$, together with their corresponding uncertainties for each fit.
\begin{table}[th!]
\centering
\caption{Values of $\chi^{2}/\mathrm{d.o.f.}$ obtained from the five fits 
corresponding to different BH-dominance criteria. Fit~1 corresponds to the most 
restrictive (1\%) BH criterion, while Fit~5 corresponds to the least restrictive (5\%) criterion.}
\label{tab:chi2}
\begin{tabular}{c @{\hspace{1.5cm}} c @{\hspace{1.5cm}} c}
\hline\hline
\textbf{Fit} & \textbf{BH criterion} & $\boldsymbol{\chi^{2} / \mathrm{d.o.f.}}$ \\
\hline
Fit 1 & $\Delta\sigma/\sigma \le 1\%$ & 0.82 \\[3pt]
Fit 2 & $\Delta\sigma/\sigma \le 2\%$ & 0.75\\[3pt]
Fit 3 & $\Delta\sigma/\sigma \le 3\%$ & 0.73 \\[3pt]
Fit 4 & $\Delta\sigma/\sigma \le 4\%$ & 0.68 \\[3pt]
Fit 5 & $\Delta\sigma/\sigma \le 5\%$ & 0.69 \\
\hline\hline
\end{tabular}
\end{table}
\begin{table}[th!]
\centering
\caption{Best-fit values of the parameters $a_E$ and $a_M$ obtained from the 
five analyses corresponding to different BH-dominance criteria. 
Quoted uncertainties represent the $68\%$ confidence level as determined 
from the Hessian method.}
\label{tab:fit_params}
\begin{tabular}{c 
        @{\hspace{1.5cm}} c 
        @{\hspace{1.5cm}} c}
\hline\hline
\textbf{Fit} & \textbf{$\boldsymbol{a_E}$ (GeV$^{2}$)} & \textbf{$\boldsymbol{a_M}$ (GeV$^{2}$)} \\
\hline
Fit 1 & 0.966 $\pm$ 0.075 & 0.585 $\pm$ 0.096 \\[3pt]
Fit 2 & 0.982 $\pm$ 0.058 & 0.531 $\pm$ 0.073 \\[3pt]
Fit 3 & 1.033 $\pm$ 0.048 & 0.476 $\pm$ 0.054 \\[3pt]
Fit 4 & 1.005 $\pm$ 0.043 & 0.516 $\pm$ 0.053 \\[3pt]
Fit 5 & 0.958 $\pm$ 0.040 & 0.562 $\pm$ 0.055 \\
\hline\hline
\end{tabular}
\end{table}

\begin{figure}[!htb]
    \centering
\includegraphics[scale=0.6]{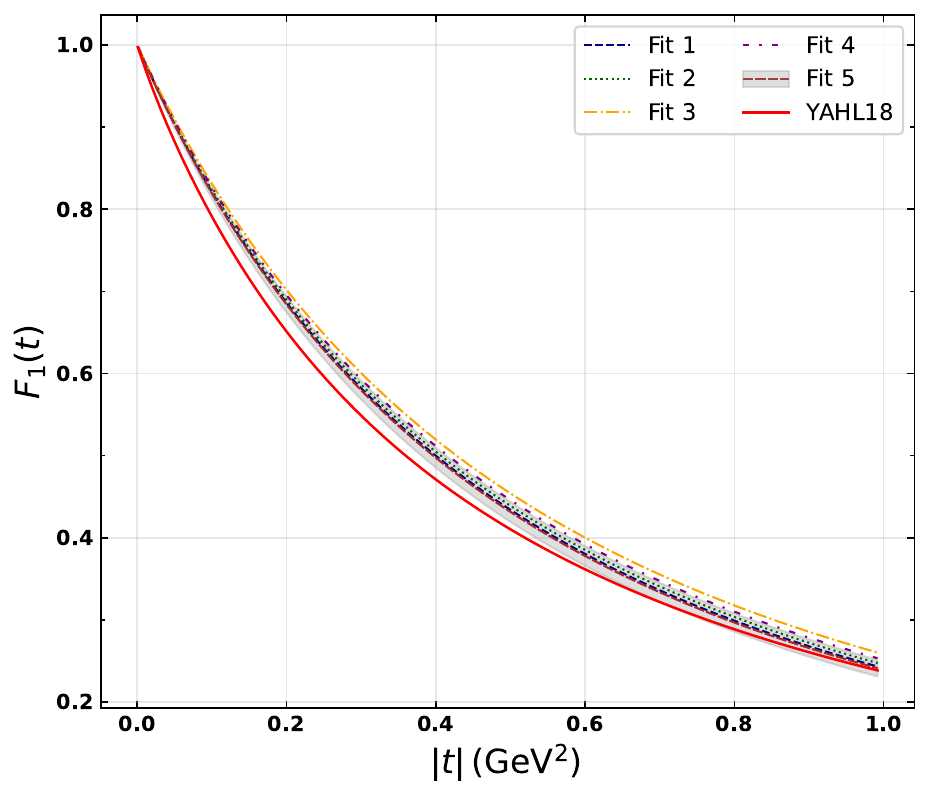}   
\caption{Comparison of the Dirac form factor $F_{1}(t)$ extracted from the five analyses performed in this work with the YAHL18 parametrization~\cite{Ye:2017gyb}. For clarity, only the uncertainty band of Fit~5 is shown, corresponding to the analysis with the largest number of data points. }
\label{fig:F1}
\end{figure}

Figure~\ref{fig:F1} compares the Dirac form factor $F_{1}(t)$ extracted from the five analyses performed in this work with the corresponding result from the YAHL18 parametrization~\cite{Ye:2017gyb}, which is based on a global fit to a large set of elastic electron-nucleon scattering data. As seen in the figure, the differences among our five fits are small once the associated uncertainties are taken into account. For clarity, only the uncertainty band of Fit~5, which contains the largest number of data points, is shown. The uncertainty bands obtained from Fits~1-4 are slightly larger but qualitatively similar in shape. In contrast, a noticeable difference appears between our BH-based extraction and the YAHL18 result: the YAHL18 curve does not fall within the Fit~5 uncertainty band. This suggests that EP measurements provide complementary and potentially important constraints on the electromagnetic FFs in the small-$|t|$ region, and therefore on derived quantities such as the nucleon electric radius. We emphasize that the fits are constrained solely by data in the range
$0.110 < |t| < 0.450~\mathrm{GeV}^2$; results shown outside this region correspond to extrapolations of the fitted parametrizations.

\begin{figure}[!htb]
    \centering
\includegraphics[scale=0.6]{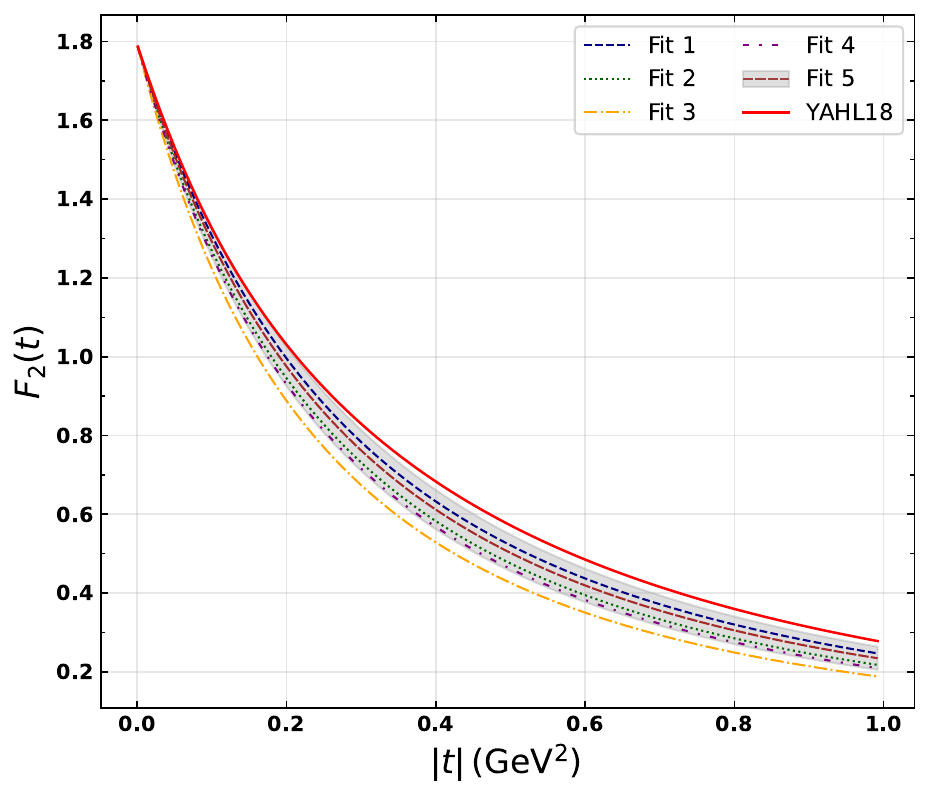}   
\caption{Same as Fig.~\ref{fig:F1}, but for the Pauli form factor $F_{2}(t)$.  }
\label{fig:F2}
\end{figure}

Figure~\ref{fig:F2} shows the same comparison as Fig.~\ref{fig:F1}, but for the Pauli form factor $F_{2}(t)$. In this case, the spread among the results of our five analyses is noticeably larger than for $F_{1}(t)$, reflecting the stronger sensitivity of $F_{2}(t)$ to the limited $|t|$ coverage of the EP data. Even more spectacular is the difference between our BH-based extraction and the YAHL18 parametrization: the YAHL18 curve lies well outside the Fit~5 uncertainty band over most of the displayed region. 
These differences are not unexpected, since $F_{2}(t)$ is more directly related to the spin-flip structure of the proton and is typically less well constrained than $F_{1}(t)$.

\begin{figure}[!htb]
    \centering
\includegraphics[scale=0.58]{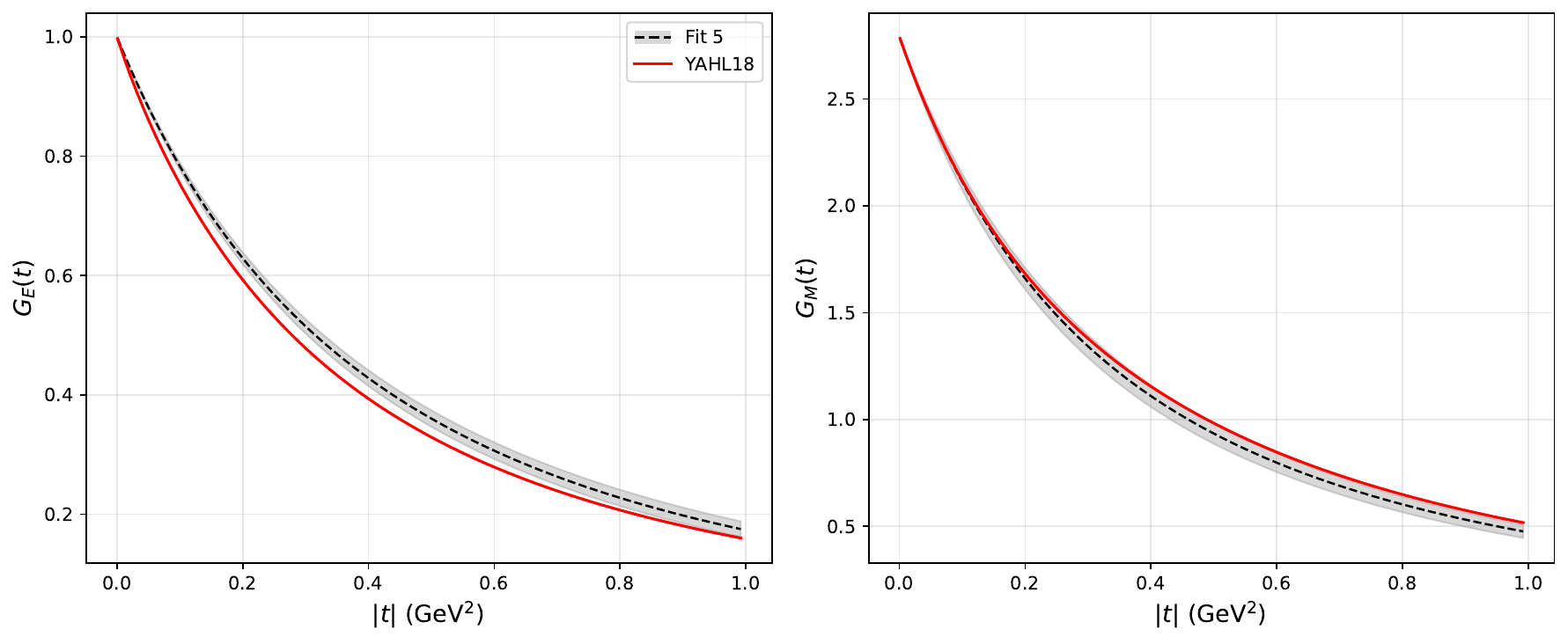}   
\caption{Comparison of the Sachs electric (left panel) and magnetic (right panel) FFs, $G_{E}(t)$ and $G_{M}(t)$, obtained from our Fit~5 analysis with the corresponding results from the YAHL18 parametrization~\cite{Ye:2017gyb}.  }
\label{fig:GE_GM}
\end{figure}

Comparing Figs.~\ref{fig:F1} and~\ref{fig:F2}, we observe a clear pattern: our BH-based extractions leads to values of $F_{1}(t)$ that are systematically larger than those obtained from YAHL18, whereas the opposite trend is observed for the Pauli form factor $F_{2}(t)$. So, it is natural to investigate the impact of these differences on the Sachs FFs, $G_{E}(t)$ and $G_{M}(t)$, defined through the well-known linear combinations
\begin{align}
\label{Eq10}
G_M(t) &= F_1(t) + F_2(t), \nonumber \\ 
G_E(t) &= F_1(t) - \tau F_2(t).
\end{align}
Figure~\ref{fig:GE_GM} shows our Fit~5 results for the electric form factor $G_{E}(t)$ (left panel) and the magnetic form factor $G_{M}(t)$ (right panel), compared with the YAHL18 parametrization. A significant deviation remains in $G_{E}(t)$ which reflects the combined effect of the differences in both $F_{1}$ and $F_{2}$. In contrast, the discrepancy is substantially reduced in case of $G_{M}(t)$, which is expected since it involves a direct sum of the two form factors. To be more precise, this reduction is due to a partial 
cancellation of $F_{1}(t)$ and  $F_{2}(t)$ respective deviations (see Figs.\ref{fig:F1} and~\ref{fig:F2}). Meanwhile, the sensitivity of $ G_E(t) $ to variations in $ F_2(t) $
grows as $ |t| $ increases because the factor $ \tau $ strongly suppresses the impact of the Pauli form factor at small $ |t| $. 
This comparison indicates how EP data can influence the extraction of the Sachs FFs and, consequently, derived quantities such as the proton charge and magnetic radii.

\subsection{$ P $-pole fit}

Now, it is of interest to examine what happens when we introduce more flexibility into the parametrizations of $F_{1}(t)$ and $F_{2}(t)$ by considering, for example, a $P$-pole form,
\begin{align}
F_1(t) &= \left(1 - \frac{t}{a_E}\right)^{-P_E}, \nonumber \\
F_2(t) &= \kappa \left(1 - \frac{t}{a_M}\right)^{-P_M}\,.
\label{Eq11}
\end{align}
However, because the EP data used in this work do not provide sufficient constraints on the Pauli form factor $F_{2}(t)$, allowing all four parameters $a_E$, $a_M$, $P_E$, and $P_M$ to vary freely leads to unphysical results. For this reason, we perform two additional fits by considering the simplifying assumptions $a_E = a_M \equiv a$ and $P_E = P_M \equiv P$. In both cases, we restrict the analysis to Set~5, which contains the largest number of data points. We call these fits as Fit~6 and Fit~7, respectively.

Both fits result in the same value of $\chi^{2}/\mathrm{d.o.f.} = 0.67$, which is slightly smaller than the value obtained for Fit~5 ($0.69$) using the dipole parametrization. The optimal fit parameters, together with their uncertainties, are listed in Table~\ref{tab:fit_params_pppole}. Introducing additional flexibility into the parametrization increases the number of free parameters and their correlations, which leads to a less constrained extraction of the form factors, particularly when extrapolated beyond the kinematic region covered by the data.
\begin{table}[t!]
\centering
\caption{Best-fit parameters and uncertainties for the $P$-pole parametrizations
defined in Eq.~(\ref{Eq11}). Fit~6 uses the condition $a_E = a_M \equiv a$,
while Fit~7 assumes $P_E = P_M \equiv P$. Both fits are performed using data Set~5.}
\label{tab:fit_params_pppole}

\begin{tabular}{c @{\hspace{1.0cm}} c @{\hspace{1.0cm}} c}
\hline\hline
\textbf{Fit} & \textbf{Parameter} & \textbf{Value} \\ 
\hline
\multirow{3}{*}{Fit~6 } 
  & $a$       & $0.534 \pm 0.066$ \\
  & $P_E $  & $1.206 \pm 0.132$ \\
  & $P_M $  & $1.960 \pm 0.215$ \\

\hline
\multirow{3}{*}{Fit~7} 
  & $a_E$  & $0.650 \pm 0.079$ \\
  & $a_M$  & $0.350 \pm 0.058$ \\  
  & $P$       & $1.417 \pm 0.142$ \\
\hline\hline
\end{tabular}
\end{table}

\subsection{Fixing $F_2(t)$}

As shown in the previous sections, the EP data considered in this analysis do not provide sufficient constraints on the Pauli form factor $F_{2}(t)$, because the data cover only the limited kinematic region $0.11 < |t| < 0.45~\mathrm{GeV}^{2}$. The sensitivity to $F_{2}(t)$ decreases as $|t|$ decreases, due to the suppressing factor $\tau = -t/(4M^{2})$ that accompanies $F_{2}(t)$ in the BH cross section (see, e.g., Eq.~(\ref{Eq7})).

This effect can be indicated by computing the BH cross section under two extreme assumptions: (i) setting $F_{1}(t)=0$ while keeping $F_{2}(t)$, and (ii) setting $F_{2}(t)=0$ while keeping $F_{1}(t)$. Comparing these reduced cross sections with the full BH result provides an estimate of the relative contributions of the Dirac and Pauli FFs in the small-$|t|$ region.

To this aim, we evaluate the BH cross section in a representative kinematic region where the BH contribution is dominant in the the full EP cross section, namely
\[
x_B = 0.1,\qquad Q^2 = 1~\mathrm{GeV}^2,\qquad \phi = 3.012,\qquad E = 6~\mathrm{GeV},
\]
and examine the dependence of the BH cross section and the isolated $F_{1}$ and $F_{2}$ contributions as functions of $|t|$ over the $|t|$-range covered by the CLAS data.

\begin{figure}[!htb]
    \centering
\includegraphics[scale=0.6]{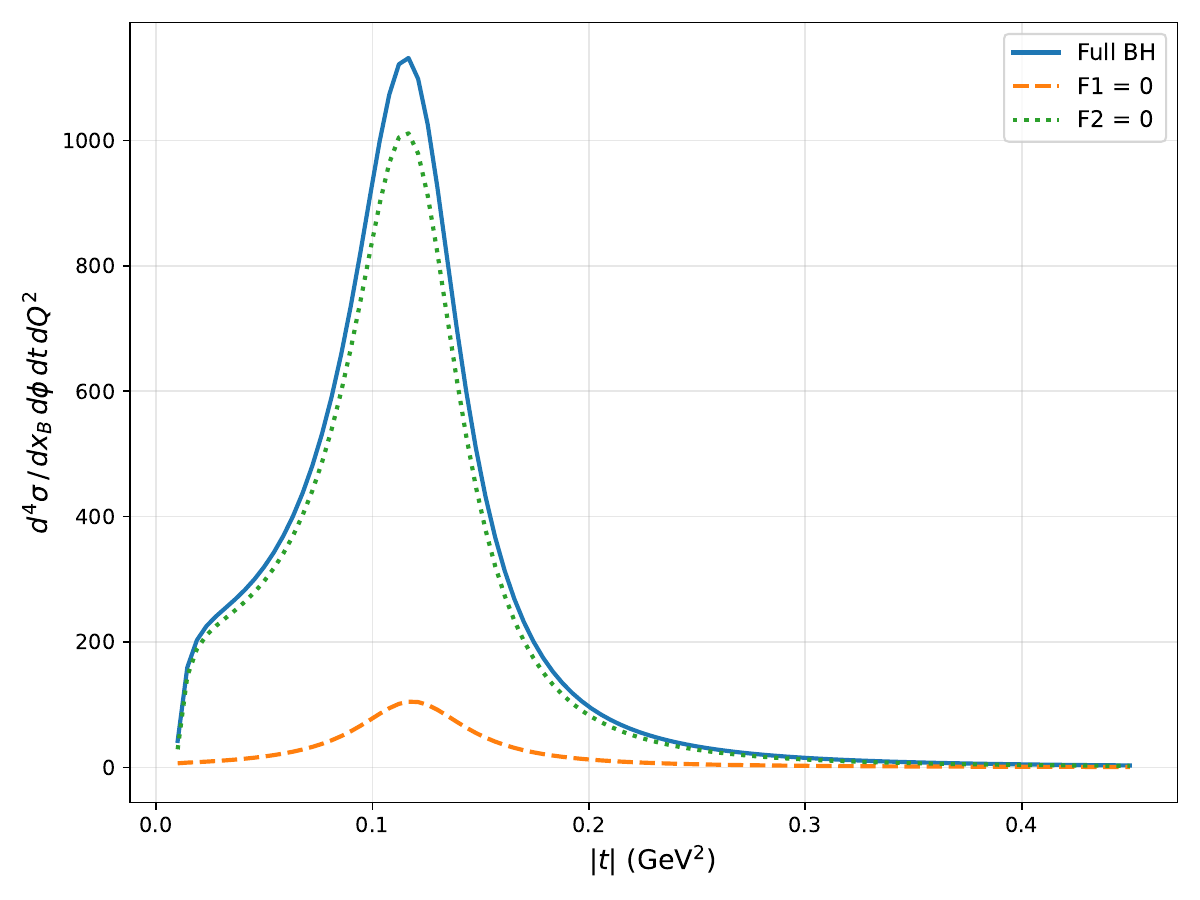}   
\caption{BH cross section and the separate contributions 
associated with the Dirac and Pauli FFs, $F_{1}(t)$ and $F_{2}(t)$, 
evaluated using the Kelly's parametrization~\cite{Kelly:2004hm}. The calculation 
is performed at a representative kinematic point where the BH process dominates 
the DVCS cross section: $x_{B}=0.1$, $Q^{2}=1~\text{GeV}^{2}$, 
$\phi = 3.012~\text{rad}$, and beam energy $E=6~\text{GeV}$.}
\label{fig:BH_F1F2}
\end{figure}

Figure~\ref{fig:BH_F1F2} shows the results obtained for the BH cross section and its individual $F_{1}(t)$ and $F_{2}(t)$ contributions, using the Kelly's parametrization~\cite{Kelly:2004hm} for both form factors. As can be seen, in this kinematic region the Dirac form factor $F_{1}(t)$ is almost totally dominant in the BH cross section. This indicates that it is very difficult to constrain $F_{2}(t)$ from the EP data included in our analysis. Consequently, it is possible that the extracted values of $F_{1}(t)$ and $F_{2}(t)$ in the previous sections may be somewhat overestimated or underestimated.

To further investigate this issue, we perform an additional analysis, referred to as Fit~8, in which $F_{2}(t)$ is fixed to the Kelly's parametrization~\cite{Kelly:2004hm}, and only $F_{1}(t)$ is fitted using a $P$-pole form. This fit leads to $\chi^{2}/\mathrm{d.o.f.} = 0.68$, with the parameters
\[
a_E = 0.565 \pm 0.100, \qquad P_E = 1.318 \pm 0.193.
\]

\begin{figure}[!htb]
    \centering
\includegraphics[scale=0.53]{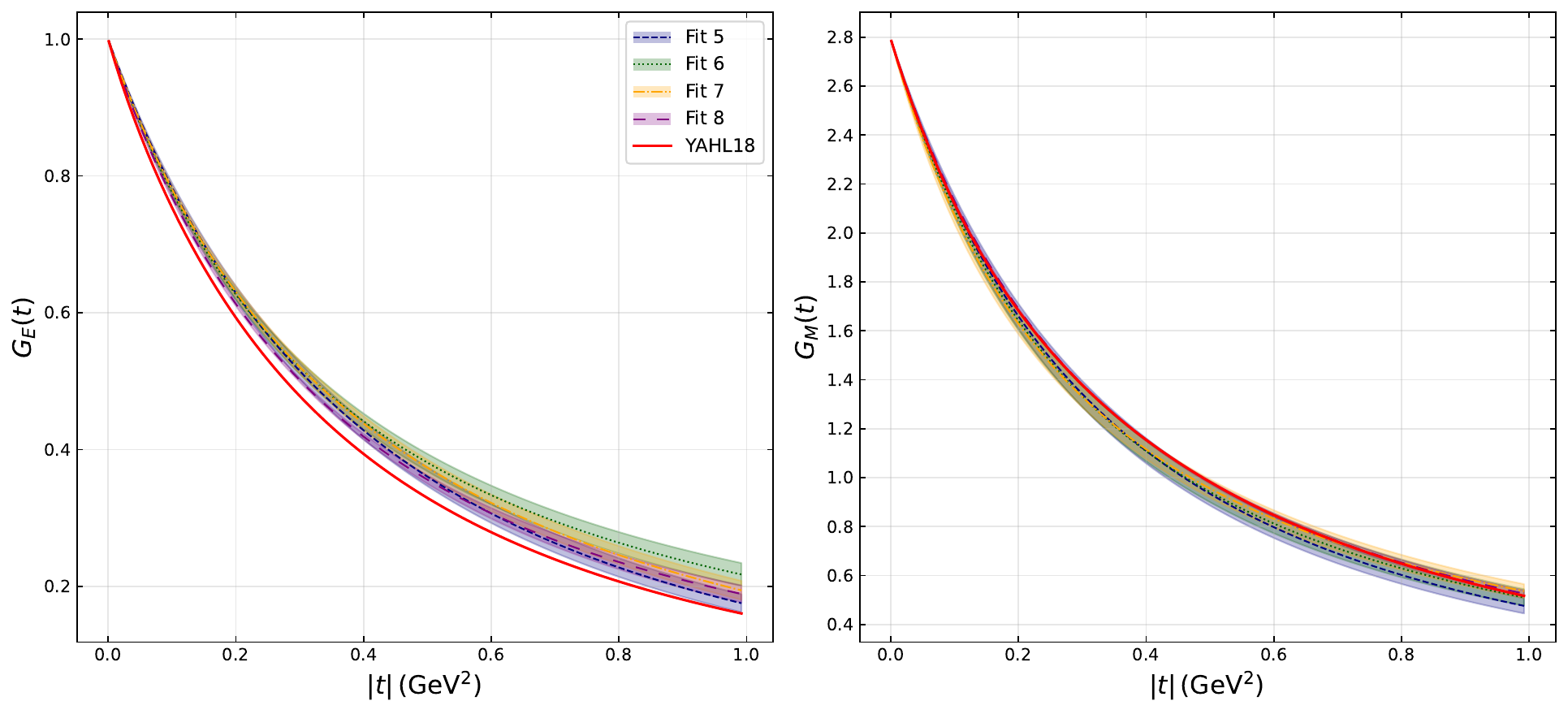}   
\caption{Comparison of the Sachs form factors $G_{E}(t)$ (left panel) and 
$G_{M}(t)$ (right panel) obtained from the different fits performed in this 
work (Fit~5, Fit~6, Fit~7, and Fit~8), together with the corresponding results 
from the YAHL18 analysis~\cite{Ye:2017gyb}.}
\label{fig:GE_GM_Final}
\end{figure}

Figure~\ref{fig:GE_GM_Final} shows a comparison of the Sachs form factors $G_{E}(t)$ and $G_{M}(t)$ obtained from the various analyses performed in this work, namely Fit~5, Fit~6, Fit~7, and Fit~8, as well as the corresponding results from the YAHL18 analysis~\cite{Ye:2017gyb}. As can be seen, the results for $G_{M}(t)$ exhibit better overall consistency than those for $G_{E}(t)$. Among our fits, Fit~8 shows the best agreement with YAHL18, particularly at smaller values of $|t|$. However, the YAHL18 curve does not fall within the uncertainty band of our results even when different parametrization forms and extraction
strategies are considered. This supports the interpretation of EP measurements as a complementary tool of constraints on the electromagnetic form factors.

Given the differences observed between the various analyses, it is now of interest to examine how these variations may affect the extracted charge and magnetic radii of the proton. Note that these quantities are highly sensitive to the small-$|t|$ behavior of the Sachs FFs, since both radii are determined by the slopes of $G_E(t)$ and $G_M(t)$ at $t = 0$.

\subsection{The charge and magnetic radii}

The charge and magnetic radii are among the most fundamental observables that provide critical insights into the electromagnetic structure of nucleons and key benchmarks for testing theoretical models of QCD~\cite{Karr:2020wgh,Pohl:2010zza,Gao:2021sml,Xiong:2023zih,Meissner:2022rsm,Cui:2022fyr,Ridwan:2023ome}. These radii describe the spatial distribution of the electric charge and magnetization within the nucleon~\cite{Chen:2023dxp}. Historically, the proton's charge radius has been a subject of intense investigations, especially in connection with the long-standing ``proton radius puzzle''~\cite{Goharipour:2025yxm}. The mean-squared charge and magnetic radii of the proton are obtained from the slopes of the Sachs FFs at $ t\rightarrow 0 $,
\begin{align}
\left<r_{E}^2\right> &= \left.  6 \dv{G_E}{t} \right|_{t=0} \,, \nonumber \\
\left<r_{M}^2\right> &= \left.  \frac{6}{\mu_p} \dv{G_M}{t} \right|_{t=0}\,,
\label{Eq12}
\end{align}
where $ \mu_p $ is the magnetic moment of the proton. 

Therefore, it is  of interest to calculate these radii using the Sachs FFs obtained from the various analyses performed in the previous sections. Table~\ref{tab:radii} reports the values of $r_{E}$ and $r_{M}$ obtained from Fits~5, 6, 7, and~8. As can be seen, there are significant differences among the results from these analyses. Fit~5, which employs a dipole parametrization for both the Dirac and Pauli FFs, leads to a remarkably small value for $r_{E}$, far below the PDG 2024 average~\cite{ParticleDataGroup:2024cfk}, $r_{E}=0.8409 \pm 0.0004$ fm. Comparing with Figs.~1 and~2 of Ref.~\cite{Goharipour:2025yxm}, this value is consistent only with a subset of lattice QCD calculations~\cite{Alexandrou:2018sjm,Alexandrou:2020aja,Jang:2019jkn}. In contrast, the magnetic radius extracted from Fit~5 agrees well with the PDG value, $r_{M}=0.851 \pm 0.026$ fm, especially when uncertainties are taken into account.
\begin{table}[t!]
\centering
\caption{Proton charge and magnetic radii extracted from the Sachs FFs obtained using Fits~5--8. The quantities $r_{E}$ and $r_{M}$ are computed from the slopes of $G_E(t)$ and $G_M(t)$ at $t=0$ according to Eq.~(\ref{Eq12}).}
\label{tab:radii}
\begin{tabular}{c
        @{\hspace{0.9cm}} c
        @{\hspace{0.9cm}} c}
\hline\hline
\textbf{Fit} & $\boldsymbol{r_E}$ \textbf{(fm)} & $\boldsymbol{r_M}$ \textbf{(fm)} \\
\hline
Fit 5 & $0.779 \pm 0.013$ & $0.842 \pm 0.027$ \\[4pt]
Fit 6 & $0.804 \pm 0.011$ & $0.860 \pm 0.022$ \\[4pt]
Fit 7 & $0.792 \pm 0.012$ & $0.889 \pm 0.030$ \\[4pt]
Fit 8 & $0.815 \pm 0.011$ & $0.831 \pm 0.004$ \\
\hline\hline
\end{tabular}
\end{table}

Fits~6 and~7, which utilize the more flexible $P$-pole parametrization, lead to larger values of $r_{E}$ than Fit~5, though still significantly smaller than the PDG value. They also predict magnetic radii larger than the PDG average, particularly for Fit~7. As discussed earlier, Fit~8 is the most reliable among our analyses, since $F_{2}(t)$ is fixed to the well-established Kelly's parametrization~\cite{Kelly:2004hm} to avoid the unconstrained behavior that arises when both FFs are simultaneously fitted. In this case as well, the extracted $r_{E}$ remains smaller than the PDG result. Taking all results obtained here together, and considering the systematic uncertainties associated with the BH-dominance assumption and
the chosen parametrizations, one can conclude that the EP data used in this study prefer smaller values of $r_{E}$ compared with those from traditional elastic electron-proton scattering.
It is worth noting that the very-low-$|t|$ measurement from the PRad experiment~\cite{Xiong:2019umf} also finds a smaller charge radius, $r_{E}=0.831 \pm 0.014$, which is compatible with our Fit~8 within uncertainties. As expected, Fit~8 also results in a more consistent value for $r_{M}$, and its very small uncertainty reflects the fact that $F_{2}(t)$ is considered fixed in this fit.

The results presented in this study indicate that measurements of DVCS observables, in kinematic regions where the BH contribution dominates, can provide a complementary tool to put new and precise constraints on the electromagnetic FFs, particularly the Dirac form factor $F_{1}(t)$. Such measurements may even suggest a different $|t|$ dependence compared with that extracted from traditional elastic electron--nucleon scattering measurements. EP data, especially when covering smaller values of $|t|$, also have the potential to provide new insights into the proton charge radius and may offer additional insight into the long-standing charge-radius puzzle. Therefore, the present analysis can serve as a guideline for extracting the proton electromagnetic FFs from EP measurements. It can be further strengthened by including additional EP measurements, as well as performing a simultaneous analysis with elastic electron-nucleon scattering data.

%
%%%%%%%%%%%%%%%%%%%%%%%%%%%%%%%%%%%%%%%%%%%%%%%%%%%%%%%%%%%%%%%%%%%%%%%%%

\section{Summary and conclusion}\label{sec:six}

In the present study, based on CLAS measurements of exclusive photon leptoproduction, we performed a comprehensive analysis of the proton electromagnetic FFs in kinematic regions where the BH process dominates the cross section. Considering the strong dependence of the BH amplitude on the Dirac and Pauli FFs, $F_{1}(t)$ and $F_{2}(t)$,
we examined several fitting scenarios based on dipole and $P$-pole parametrizations and extracted them as well as the corresponding Sachs FFs, $G_{E}(t)$ and $G_{M}(t)$.

Our results indicate that the EP data in the range $0.11 < |t| < 0.45~\text{GeV}^{2}$ can provide constraints on the Dirac form factor $F_{1}(t)$, whereas the sensitivity to the Pauli form factor $F_{2}(t)$ remains limited due to the reduced impact of its contribution at small values of $|t|$. 

From the slopes of $G_{E}(t)$ and $G_{M}(t)$ at $t=0$, we determined the proton charge and magnetic radii. All fits consistently indicate smaller charge-radius values compared with those obtained from traditional elastic electron-proton scattering measurements and the PDG average results. However, Fit~8, in which $F_2(t)$ was fixed to the Kelly's parametrization, provides the most reliable determination and shows  qualitative agreement with the PRad result within uncertainties.  In contrast, the magnetic radius, exhibits good consistency with the PDG value across different fit scenarios.

Overall, the present study indicates that EP process, when measured in regions dominated by the BH process, provide a complementary tool for constraining the electromagnetic structure of the proton. Extending EP measurements to even smaller values of $|t|$ allow for a more precise determination of the nucleon radii and may offer additional insight into the long-standing charge-radius puzzle. The analysis framework developed in this study provides a solid foundation for future combined studies including both EP and elastic electron-nucleon scattering data which enables a unified  extraction of the nucleon electromagnetic FFs.

In addition, the methodology developed in this work can be directly applied to other exclusive photon leptoproduction measurements, such as those from Hall~A~\cite{JeffersonLabHallA:2015dwe,Defurne:2017paw}, provided that sufficiently precise cross-section data are available in kinematic regions where the BH contribution is dominant. This offers promising opportunities for future studies with expanded kinematic coverage and improved precision.

%
%%%%%%%%%%%%%%%%%%%%%%%%%%%%%%%%%%%%%%%%%%%%%%%%%%%%%%%%%%%%%%%%%%%%%%%%% 
\section*{ACKNOWLEDGEMENTS}

We thank K. Passek-K. for her significant advisory role and valuable comments. We also thank K. Kumerički for providing a sample notebook for using the \texttt{Gepard} package. M. Goharipour thanks the Theoretical Physics Department at CERN for kindly hosting him during part of this project.

%
%%%%%%%%%%%%%%%%%%%%%%%%%%%%%%%%%%%%%%%%%%%%%%%%%%%%%%%%%%%%%%%%%%%%%%%%% 
%\section*{Note Added}
%We are pleased to provide  the GPDs extracted in the present study with their uncertainties in any desired values of $ x $ and $ -t $ upon request.

%
%%%%%%%%%%%%%%%%%%%%%%%%%%%%%%%%%%%%%%%%%%%%%%%%%%%%%%%%%%%%%%%%%%%%%%%%% 

\bibliographystyle{apsrev4-1}
\bibliography{article}

\end{document}